\documentclass[twocolumn,amsmath,amssymb]{revtex4}
\usepackage{amsmath,amsthm,amssymb}
\input epsf.tex
\usepackage{latexsym}
\usepackage[dvips]{graphicx}

\begin{document}

\newcommand{\comment}[1]{{\bf\small [#1]} }

\def\lsim{\mathrel{\rlap{\lower4pt\hbox{\hskip1pt$\sim$}}
    \raise1pt\hbox{$<$}}}
\def\gsim{\mathrel{\rlap{\lower4pt\hbox{\hskip1pt$\sim$}}
    \raise1pt\hbox{$>$}}} 
\newcommand{\vev}[1]{ \left\langle {#1} \right\rangle }
\newcommand{\bra}[1]{ \langle {#1} | }
\newcommand{\ket}[1]{ | {#1} \rangle }
\newcommand{\ev}{ {\rm eV} }
\newcommand{\kev}{{\rm keV}}
\newcommand{\mev}{{\rm MeV}}
\newcommand{\gev}{{\rm GeV}}
\newcommand{\tev}{{\rm TeV}}
\newcommand{\mpl}{$M_{Pl}$}
\newcommand{\mw}{$M_{W}$}
\newcommand{\Ft}{F_{T}}
\newcommand{\Zparity}{\mathbb{Z}_2}
\newcommand{\BLambda}{\boldsymbol{\lambda}}
\newcommand{\be}{\begin{eqnarray}}
\newcommand{\ee}{\end{eqnarray}}
\newcommand{\cl}{95\% C.L.}

\def\rd{\mathrm{d}}
\def\lep{{\sc lep} }
\def\lepone{{\sc lep} 1 }
\def\leptwo{{\sc lep} 2 }
\def\aleph{{\sc aleph} }
\def\opal{{\sc opal} }
\def\delphi{ {\sc delphi} }
\newcommand{\ttbar}{$t \bar{t} $ }

\title{Top-tagging: A Method for Identifying Boosted Hadronic Tops}
\author{David E. Kaplan, Keith Rehermann, Matthew D. Schwartz and Brock Tweedie}
\affiliation{Department of Physics and Astronomy
Johns Hopkins University
Baltimore, MD 21218, U.S.A.}

\begin{abstract}
A method is introduced for distinguishing top jets (boosted, hadronically decaying top quarks) from light quark and gluon jets using
jet substructure. The procedure involves parsing the jet cluster to resolve its subjets, 
and then imposing kinematic constraints.
With this method, light quark or gluon jets with $p_T \simeq 1$ TeV
can be rejected with an efficiency of around 99\% while retaining up to 40\% of top jets.
This reduces the dijet background to heavy \ttbar resonances by a factor of $\sim\!10,000$, 
thereby allowing
resonance searches in \ttbar to be extended into the all-hadronic channel.
In addition, top-tagging can be used in \ttbar events
when one of the tops decays semi-leptonically, in events with missing energy,
and in studies of $b$-tagging efficiency at high $p_T$.
\end{abstract}
\maketitle

The Large Hadron Collider (LHC) is a top factory. 
The millions of top quarks it produces
will provide profound insights into the standard model and
its possible extensions.
Most of the tops will be produced near threshold, 
and can be identified using the same
kinds of techniques applied at the Tevatron -- looking for the
 presence of a bottom quark through $b$-tagging, identifying the $W$ boson, or finding three
jets whose invariant mass is near $m_t$. 
However, some of the top quarks produced at the LHC will be highly boosted.
In particular, almost every new physics scenario that addresses the hierarchy problem
will include new heavy particles which decay to tops (such as KK gluons in Randall-Sundrum models,
squarks in supersymmetry, top primes in little Higgs models, etc.).  
If their masses are even a factor of a few above the top mass, the tops that they produce will decay to collimated
collections of particles that look like single jets.
In this case, the standard top identification techniques may falter:
 $b$-tagging is difficult because the tracks are crowded and unresolvable,
the $W$ decay products are not always isolated from each other or from the $b$ jet,
 and the top jet mass may differ
 from $m_t$ due to an increased amount of QCD radiation. 
%Thus without
% new ideas, it may be difficult to find these boosted tops. 

In most studies of \ttbar resonances, emphasis is placed on the channel in which
one top decays semi-leptonically (to an electron or muon, a neutrino, and a $b$ jet) 
and the other hadronically~\cite{Agashe:2006hk,Baur:2008uv}. This avoids having to confront the large dijet background
to all-hadronic \ttbar. However, these studies need to assume that the lepton can be isolated,
which often excludes the electron channel, and that at least one $b$ jet is tagged, which is
difficult at high $p_T$~\cite{ATLbtag}. The hard muon tag alone already discards 90\% of the \ttbar events. So one
would like to be able to use the all-hadronic channel without $b$-tags.
In this paper, we introduce a practical and efficient method for tagging boosted hadronically-decaying tops.

\begin{figure}[t]
\resizebox{\hsize}{!}{\includegraphics{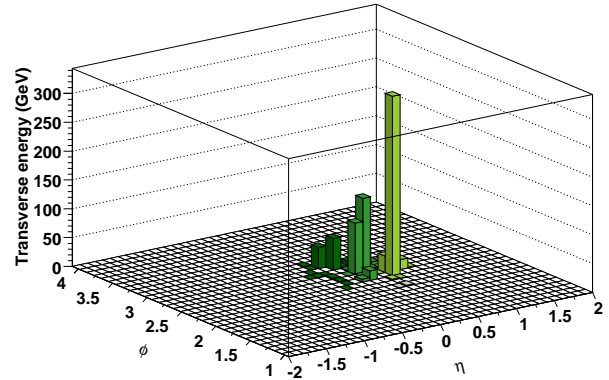}}
\caption{A typical top jet with a $p_T$ of 800 GeV at the LHC. 
The three subjets after top-tagging are shaded separately.}
\label{fig:lego}
\end{figure}

A top quark's dominant decay mode is to a $b$ quark and a $W$ boson with the
$W$ subsequently decaying to two light quarks.  The three quarks normally 
appear as jets in the calorimeter, but for highly boosted tops these jets 
may lie close together and may not always be independently resolved.  For example, a zoomed-in 
lego plot of a typical top jet is shown in Figure~\ref{fig:lego}.  
It displays energy deposited in an ideal calorimeter versus pseudorapidity, $\eta$,
and azimuthal angle, $\phi$.
The three quark jets show up clearly by eye, but it is easy 
to see how the number of jets identified by conventional clustering would be highly variable and
strongly dependent on the jet-resolution parameter.
This is the inherent difficulty with extrapolating the techniques that
work for slower tops, where the decay products are widely separated, to the boosted case.

The natural direction for finding boosted tops is to look into subjet analysis
and other measures of the energy distribution in the events.
A recent ATLAS note~\cite{Brooijmans} explored the possibility
by cutting on the jet mass and the $y_{\mathrm cut}$ variables associated with the $k_T$ algorithm.
They achieved an efficiency of 45\% for top-tagging at $p_T=1$ TeV with 1 in 20 background jets getting through.
Such efficiencies are not strong enough to filter \ttbar events from the enormous dijet background~\footnote{
While this paper was in preparation, Ref.~\cite{Thaler:2008ju} appeared. They
use soft singularities of the parton shower to distinguish
tops from background. The resulting efficiencies are similar to those of~\cite{Brooijmans}.}.

The key to efficient top-tagging is in isolating features of QCD which control the background from features particular
to the top quark. As can be seen in Figure~\ref{fig:lego}, boosted top events look like single jets with three resolvable
subjets in a small region of the calorimeter. 
These subjets are separated by angular scales
of order $\sim 2 m_t/p_T$, and so remain distinguishable from one another up to $p_T$'s of roughly 2 TeV 
for a calorimeter cell size of $0.1$.  In QCD, on the other hand, 
a typical high-$p_T$ jet starts as a single 
hard parton, which subsequently cascades into a high multiplicity of soft and collinear particles.
Most of these particles cannot be resolved by the real calorimeter, as they tend to fall into a single cell or a set of adjacent cells.
In order to look like a decayed top quark, a hard parton must at least undergo two branchings
 at somewhat large angles and energy sharings, which is relatively rare, as we will see.  
The primary task, then, is to isolate events with three hard, nearby subjets.  Subsequently, we may
exploit the full 3-body kinematics of top decay to construct additional discriminating variables. 

In order to avoid the pitfalls mentioned above for fixed-size jet clustering, we first cluster an event using a large jet radius to capture all
of the potential substructure, and then iteratively
decluster each jet to search for subjets.  Similar ideas have been employed by 
by Butterworth et al. to extract substructure in Higgs jets~\cite{Butterworth:2008iy}
and $W$ jets~\cite{Butterworth:2002tt,Butterworth:2007ke}, and
part of our algorithm is an adaptation of their method.

The top-tagging algorithm is as follows:
\begin{itemize}
	\item First, particles are clustered into jets of size $R$. For this step, we use the Cambridge-Aachen (CA)
algorithm~\cite{Dokshitzer:1997in,Wobisch:1998wt}. This iterative procedure begins with all four-vectors in an event, 
as defined by the energy deposits in the calorimeter. It then finds the pair which is closest in 
$\Delta R = \sqrt{\Delta\eta^2 + \Delta \phi^2}$,
merges it into a single four-vector, and then repeats. 
The procedure ends when no two four-vectors have $\Delta R < R$.
	\item Next, each jet in the event (for \ttbar this would be one of the hardest two) is 
declustered, to look
for subjets. This is done by reversing each step in the CA clustering, iteratively separating each jet into two objects.
The softer of the two objects is thrown out if
its $p_T$ divided by the full jet $p_T$ is less than a parameter $\delta_p$, 
and the declustering continues on the harder object.
	\item The declustering step is repeated until one of four things happens:
1) both objects are harder than $\delta_p$; 
2) both objects are softer than $\delta_p$;
3) the two objects are too close, $|\Delta \eta| + |\Delta \phi| < \delta_r$,
where $\delta_r$ is an additional parameter; or
4) there is only one calorimeter cell left.
In case 1), the two hard objects are considered subjets. 
In cases 2), 3), and 4), the original jet is considered irreducible. 
	\item If an original jet declusters into two subjets,
the previous step is repeated on those subjets 
(with $\delta_p$ still defined with respect to the original jet's $p_T$)
resulting in 2, 3, or 4 subjets of the original jet.
The cases with 3 or 4 subjets are kept, the 4th representing
an additional soft gluon emission, while the 2 subjet case is rejected.
	\item With these 3 or 4 subjets in hand, additional kinematic cuts are imposed:
the total invariant mass should be near $m_t$, two subjets should reconstruct $m_W$,
and the $W$ helicity angle should be consistent with a top decay, as described below.
\end{itemize}

For our particular implementation, 
we simulate dijet events and \ttbar events in the standard model 
at the LHC using {\sc pythia} {\tt v.6.415}~\cite{Sjostrand:2006za}. 
In order to simulate the resolution of the ATLAS or CMS calorimeters, 
particles in each event are combined into square bins of 
size $\Delta \eta = \Delta \phi = 0.1$, which are interpreted as massless four-vector ``particles'' and
inputted into the clustering routine.
For jet clustering, we employ the CA algorithm as implemented in {\sc fastjet} 
{\tt v.2.3.1}~\cite{Cacciari:2005hq}. 
Because more highly boosted tops will be more collimated, 
we correlate the jet clustering parameter $R$, the event's scalar $E_T$, and the two clustering
parameters $\delta_p$ and $\delta_r$ as follows:
for $E_T > 1000,1600,2600$ GeV, we take $R=0.8,0.6,0.4$, $\delta_p=0.10,0.05,0.05$ and 
$\delta_r = 0.19,0.19,0.19$ respectively.
Then we demand that the jets be hard by putting a cut on the jet $p_T$ scaled by
the event's scalar $E_T$: $p_T > 0.7\frac{E_T}{2}$. 
Both jets must also satisfy the absolute constraints $p_T>500$ GeV and $|\eta| < 2.5$ to
be considered for analysis. 

Next, we perform the subjet decomposition, demanding 3 or 4 subjets, as described above.
For jets with  $p_T<1000$ GeV, we then ask that the invariant mass of the sum of the subjet four-vectors be within 30 GeV of the top mass (145-205 GeV)
and that there exist two subjets which reconstruct the $W$ mass to within 15 GeV (65-95 GeV). 
Harder jets will have broader mass distributions, due to increased radiation from QCD. Thus,
if a jet has $p_T>1000$ GeV, we shift the upper ranges of top and $W$ mass cuts to 
$p_T/20+ 155$  GeV and 
$p_T/40 + 70$ GeV
respectively.
Finally, we demand that the $W$ helicity angle satisfy $\cos\theta_h < 0.7$, as we now explain. 

\begin{figure}[t]
\resizebox{\hsize}{!}{\includegraphics{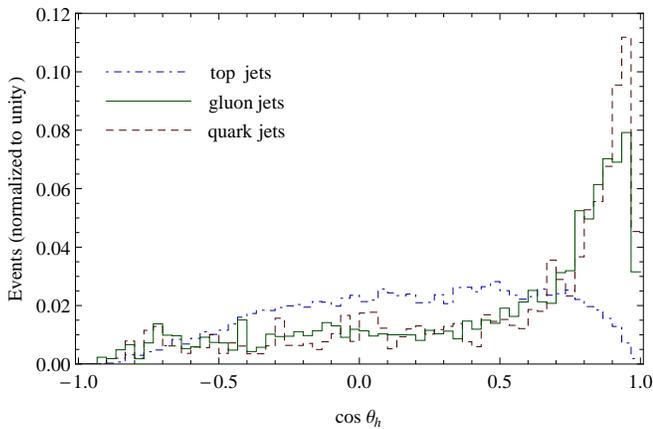}}
\caption{Distribution of helicity angle for top jets, gluon jets, and light quark jets
for $p_T > 700$ GeV. These distributions are after the subjet requirement, top mass cut, and $W$ mass cut have been imposed.}
\label{fig:hel}
\end{figure}

The helicity angle is a standard observable in top decays, used to
determine the Lorentz structure of the top-$W$ coupling~\cite{Chwalek:2007pc}.
It is defined
as the angle, measured in the rest frame of the reconstructed $W$, between
the reconstructed top's flight direction and one of the $W$ decay products.
Normally, it is studied in semi-leptonic top decays, where the charge of
the lepton uniquely identifies these decay products.  In hadronic top
decays there is an ambiguity which we resolve by choosing the lower $p_T$
subjet, as measured in the lab frame. (Other choices are possible
and make little difference on the final efficiencies.)  For top jets, 
the distribution is basically flat: since the $W$ decays on-shell, its decay products are almost isotropically distributed
in the $W$ rest frame. In contrast, for light quark or gluon jets, 
the distribution diverges (at the parton level) as 1/($1-\cos\theta_h$). This
corresponds to a soft singularity in the QCD matrix elements for emitting an additional parton.
Example distributions are shown in Figure~\ref{fig:hel}. The qualitative features we understand
analytically at the parton level are clearly visible after showering and hadronization.
Other observables sensitive to the soft singularity are possible~\cite{Thaler:2008ju}, and will give similar signal/background
enhancements. 

\begin{figure}[t]
\resizebox{\hsize}{!}{\includegraphics{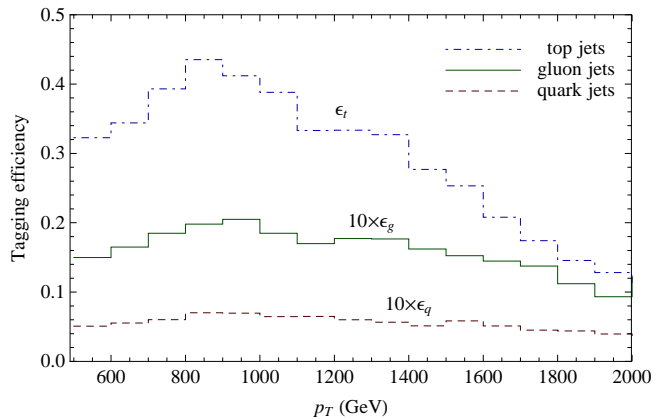}}
\caption{The efficiencies for correctly tagging a top jet ($\epsilon_t$), and mistagging a gluon jet ($\epsilon_g$) or light quark jet 
($\epsilon_q$). The quark and gluon efficiences are of order $1\%$ and have been scaled in the plot by a factor of 10 for clarity.
} 
\label{fig:eff}
\end{figure}

To check the efficacy of this method, 
we calculate the efficiency for correctly tagging a top jet, $\epsilon_t$, and the
efficiencies for mistagging light-quark or gluon jets as tops,
$\epsilon_{q}$ and $\epsilon_{g}$ respectively. These are shown in Figure~\ref{fig:eff}.
There are a few important qualitative observations one can make about this plot.
For very large $p_T$ the top-tagging efficiency goes
down. This is because these jets are so highly boosted that 
the calorimeter can no longer distinguish the subjets. As $p_T$ goes below 900 GeV, the
top-tagging efficiency also decreases. This is due to some of the top jets becoming too fat
for the initial $R=0.8$ clustering.  (This somewhat tight choice was made to suppress the mistag efficiency, which
grows faster than the top-tag efficiency with increasing $R$.)
Examples of the sequential effects of the individual cuts are shown in Table \ref{tab:cuts}.
The clustering $R$'s and kinematic cuts can be varied to increase the tagging and mistagging efficiencies, as
desired for a particular $S/\sqrt{B}$ goal.

\begin{table}[h]  \centering  \begin{tabular}{@{}|c| c|c|c|c|c| @{}}   
 \hline & $p_T$ (GeV) & subjets & $m_t$ & $m_W$ & $\theta_h$ \\ 
\hline
&    500-600 &  0.56  & 0.43  & 0.38 & 0.32  \\  
$\epsilon_t$ 
&   1000-1100 & 0.66  & 0.52 & 0.44 & 0.39  \\  
&   1500-1600 & 0.40 & 0.33  & 0.28 & 0.25  \\    
\hline   
%  & subjet & $m_t$ & $m_W$ & $\cos{\theta_h}$ \\     
\hline    
&500-600 & 0.135 & 0.045 & 0.027 & 0.015 \\     
$\epsilon_{g}$ 
&1000-1100 & 0.146 & 0.054 & 0.032 & 0.018 \\     
&1500-1600 & 0.083 & 0.038 & 0.025 & 0.015 \\    
\hline 
\hline    
&500-600 & 0.053 & 0.018 & 0.011 & 0.005 \\     
$\epsilon_{q}$ 
&1000-1100 & 0.063 & 0.023 & 0.013 & 0.006 \\     
&1500-1600 & 0.032& 0.015 & 0.010 & 0.006 \\    
\hline 
 \end{tabular} 
 \caption{Incremental efficiencies for top, gluon, and light quark jets passing the subjets, 
invariant mass, and helicity angle cuts for jets in three different $p_T$ windows.}
 \label{tab:cuts}
\end{table}

One important concern is whether the Monte Carlo generates the \ttbar and dijet distributions correctly. 
Jet substructure in particular is strongly dependent on aspects of the parton shower (both initial state and final state radiation), 
the underlying event, and the model of hadronization. To approach these issues, we redid our analysis using 
samples generated with various shower parameters, with the  ``new'' $p_T$-ordered dipole shower in {\sc pythia}, 
and with {\sc herwig} {\tt v.6.510} \cite{Corcella:2000bw}. We find a 50\% variation in $\epsilon_q$ and $\epsilon_g$
 and a negligible change in $\epsilon_t$.  We also ran {\sc pythia} with multiple interactions and initial state radiation turned off,
individually and together.  Effects on  $\epsilon_q$ and $\epsilon_g$ are at the 10\% level or less, indicating that the QCD jet substructure
relevant for top-tagging is mostly controlled by final state parton branchings.

One might also be worried about whether, since we are looking
at multi-(sub)jet backgrounds, it would be important to include full matrix element calculations. 
However, since the events are essentially two jet events, the substructure is due almost entirely 
to collinear radiation, which the parton shower should correctly reproduce~\cite{Norrbin:2000uu}. 
To confirm this, we have also simulated background
events using {\sc madgraph} {\tt v.4.2.4}~\cite{Alwall:2007st}. Using events with $2 \rightarrow 4$ matrix elements
in a region of phase space where 1 parton recoils against 3 relatively collinear partons,  we
repeated our analysis without showering or hadronization. The resulting mistag efficiencies
were consistent with those from the {\sc pythia} study to within 10\%, which provides
justification for both the parton shower approximation and the robustness of our algorithm.

One possible way to verify the Monte Carlo predictions for jet substructure would be to use data directly.
Although boosted tops are not produced at the Tevatron, there are plenty of hard dijet events.
These could be used to test the mistag efficiency, tune the Monte Carlo,
and optimize jet-tagging parameters for the LHC. In addition, at the LHC,
the efficiency of the top-tagging algorithm can be calibrated by comparing
the rate for \ttbar events where one top decays semi-leptonically with
the rate in the all-hadronic channel. The background rejection efficiency can also be studied
by looking in side-bands where the jet invariant mass is not close to $m_t$.

\begin{figure}[t]
\resizebox{\hsize}{!}{\includegraphics{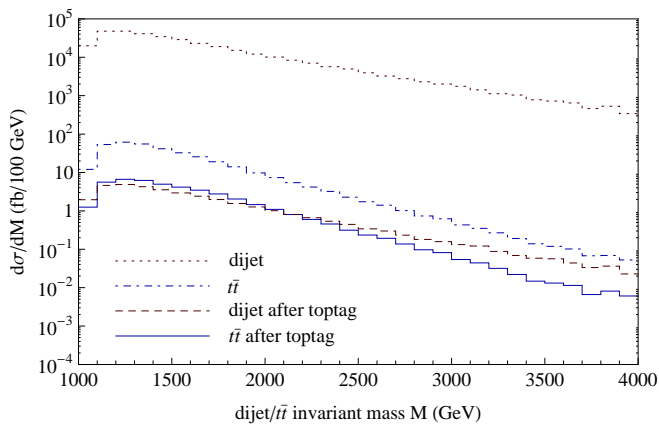}}
\caption{Effect of top jet tag on standard-model \ttbar and dijet distributions at the LHC.
Both the $t$ and $\bar{t}$ decay hadronically, and no $b$-tagging is used. With top-tagging, a
strongly-produced \ttbar resonance (not shown) would stand out clearly 
over background in this channel.}
\label{fig:massscan}
\end{figure}

Top-tagging may be particularly useful in the search for
new physics in \ttbar resonances. In the all-hadronic channel, the 
biggest background for \ttbar is dijets, so in Figure~\ref{fig:massscan} 
we show the dijet and \ttbar invariant mass distributions before and after top-tagging both jets.
It is evident that after top-tagging, the dijet sample is reduced to the level of the
\ttbar sample. As an example application, in certain Randall-Sundrum 
models~\cite{Randall:1999ee,Agashe:2004rs}
KK gluons decay dominantly to $t \bar{t}$.
It has been shown that if one can isolate the \ttbar events, the resonance will stand out
as a clean peak over the standard model \ttbar background~\cite{Agashe:2006hk,Lillie:2007yh,Baur:2008uv}.
Since top-tagging can reduce the dijet background to the size of the \ttbar background, \ttbar resonance searches can be done in the all-hadronic channel  for resonances up to a few TeV.

There are many applications for top-tagging besides \ttbar resonances searches.
For example, a common new physics signal is \ttbar pairs in association with missing 
energy~\cite{Han:2008gy}.
This may happen, for instance, in supersymmetry when heavy top squark pairs decay to
highly boosted tops and neutralinos.
Top-tagging can not only reduce the standard model backgrounds in this context, but it can also help distinguish top jets from light quark jets
in any signal event, which may be helpful in studying the flavor structure of the new physics.
In addition, top-tagging could potentially 
be applied in searches for single top events where exactly one top jet is required.
Finally, our technique could be used as a
 handle for measuring $b$-tagging efficiency at high $p_T$.

In conclusion, we have demonstrated that it is possible to distinguish highly energetic top quarks from standard model backgrounds at the LHC.
With efficiencies $\epsilon_t \! \sim \! 40\%$ and $\epsilon_q \! \sim \! \epsilon_g \! \sim\! 1\%$, top-tagging is better than $b$-tagging at high $p_T$.
Top jets can now be considered standard objects for event analysis at the LHC, as $b$ jets are at the Tevatron.

The authors would like to thank Gavin Salam and Morris Swartz for helpful conversations. 
This work is supported in part by the National Science Foundation under grant NSF-PHY-0401513, the Department of Energy's OJI program under grant DE-FG02-03ER4127, and the Johns Hopkins Theoretical Interdisciplinary Physics and Astronomy Center,
the Leon Madansky Fellowship,
 and the LHC Theory Initiative program.

\end{document}